\documentclass{aa}  

\usepackage{natbib}
\usepackage{graphicx}
\usepackage{txfonts}
\usepackage[colorlinks=true, allcolors=blue]{hyperref}
\usepackage[normalem]{ulem}
\usepackage{amsmath}
\usepackage{xcolor}
\definecolor{ochre}{rgb}{0.8, 0.47, 0.13}

\begin{document} 

   \title{The impact of natal kicks on black hole binaries}

   \author{Alejandro Vigna-G\'omez
          \inst{1}
          }

   \institute{Max-Planck-Institut f\"ur Astrophysik, 
Karl-Schwarzschild-Str. 1, 85748 Garching, Germany
             }

   \date{Received ... ; accepted ...}
 
  \abstract
    {
    In massive binary-star systems, supernova (SN) explosions can significantly alter the orbit during the formation of compact objects. 
    Some compact objects are predicted to form via direct collapse, with negligible mass loss and no baryonic ejecta emitted.
    In this scenario, most of the energy is released via neutrinos and any resulting natal kick arises from asymmetries in their emission.
    Here, I investigate stellar collapse leading to binary black hole (BH) formation, with a focus on how the natal kick influences the gravitational-wave driven merger time.
    In summary, I find that when mass loss is negligible, natal kicks are unlikely to affect binary BH merger rates, delay times, or spin orientations.
    More specifically, I find the following. For low natal kicks, the effect on the time to coalescence is negligible.
    For moderate natal kicks, if the binary remains bound, up to 50\% of binary BHs experience a decrease in their time to coalescence by more than one order of magnitude.
    For large natal kicks, while most binaries become unbound, those that remain bound may acquire retrograde orbits and/or lead to shorter coalescence times.
    For binary BH mergers, large natal kicks ($\gtrsim100$ km/s) are hard to reconcile with both neutrino natal kicks and the complete collapse scenario. 
    This suggests that retrograde orbits and shortened merger times could only arise in volatile BH formation scenarios or if spin-axis tossing is at work.
    Consequently, electromagnetic observations of BHs in massive star binaries within the Local Group offer a more effective means to probe the physics behind complete collapse. 
    Another promising population for deciphering the complete collapse scenario is that of massive, wide binaries. 
    Although Gaia may help shed light on these systems, longer observational baselines will likely be needed to fully understand the roles of neutrino natal kicks and stellar collapse in BH formation.
   }
   \keywords{Stars: black holes --
            (Stars:) binaries: general --
            (Stars:) supernovae: general
               }

   \maketitle
%
%-------------------------------------------------------------------
\section{Natal kicks in binary systems}
The role of supernova (SN) explosions on the orbits of binary systems formed in isolation has been extensively studied in various contexts, including X-ray binaries \citep[e.g.][]{1995MNRAS.274..461B,1996ApJ...471..352K}, disrupted systems resulting in runaway stars \citep[e.g.][]{1998A&A...330.1047T}, or isolated compact objects---such as neutron stars \citep[e.g.][]{2002ApJ...573..283P} and BHs \citep[e.g.][]{2022ApJ...930..159A}, and double compact object mergers \citep[e.g.][]{1996ApJ...471..352K,2000ApJ...541..319K,2017ApJ...846..170T,2024ApJ...966...17B}.

These studies assume that the SN explosion occurs instantaneously, resulting in a mass decrement in the system and a natal kick imparted to the newly formed compact object, both linked to emissions and asymmetries from baryonic matter, neutrinos, and (to a lesser extent) gravitational waves \citep[see e.g.][and references therein]{2024PhRvL.132s1403V}.

\begin{figure}
    \centering
    % [trim={left bottom right top},clip]
    \includegraphics[trim={24cm 11cm 20cm 12cm},clip,width=\columnwidth]{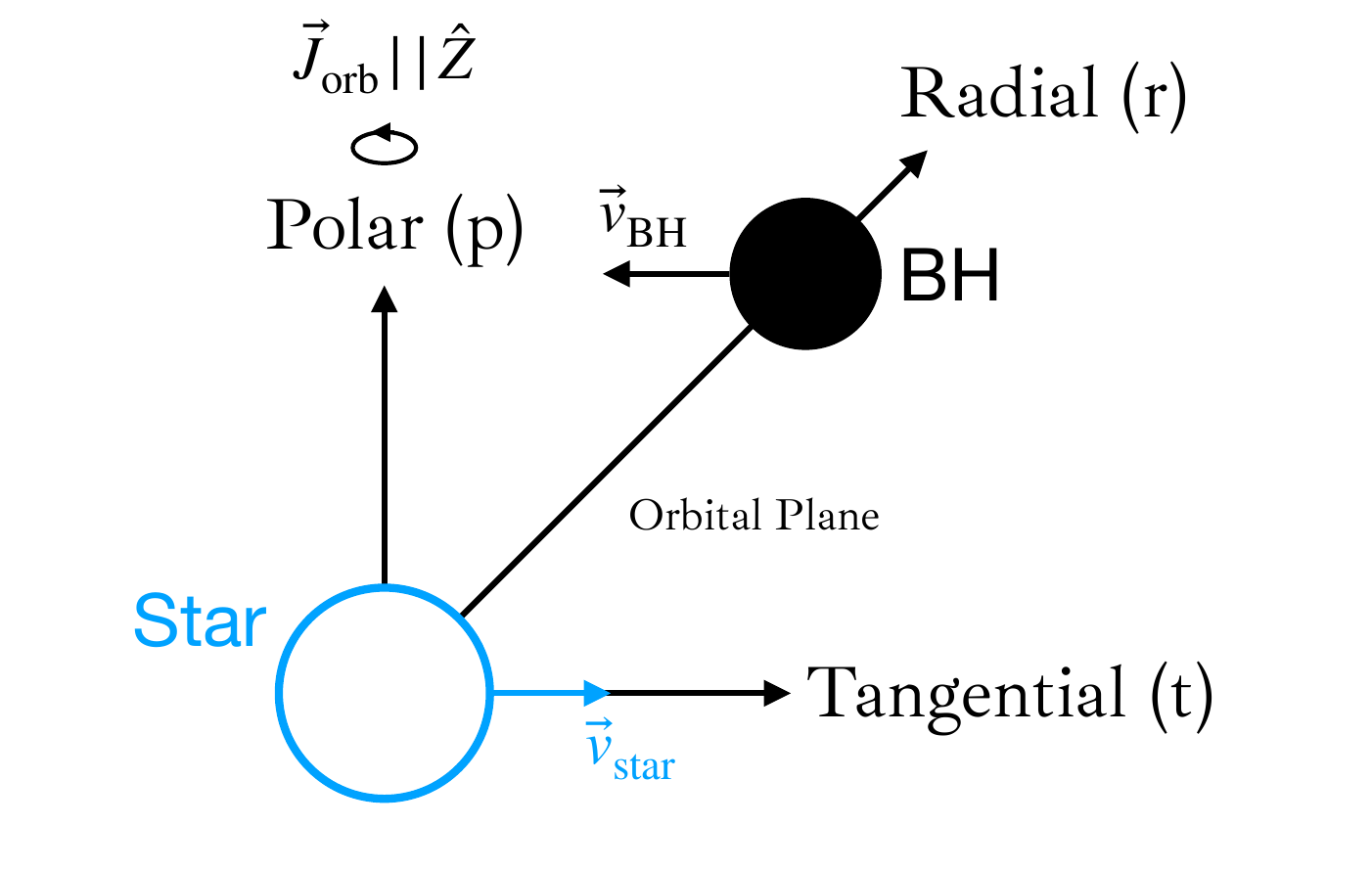}    
    \caption{
    Geometry of a circular binary system.
    The system is comprised of a star, which is destined to experience complete collapse into a BH, and its companion. In this case, the object is represented as a BH.
    The reference frame is centred on the star at the moment of collapse, and its defined by the radial (r), tangential (t), and polar (p) orthogonal axes.
    In this frame, the orbital plane is defined as the r–t plane, with the orbital angular momentum vector ($\vec{J}_{\rm orb}$) along the p–axis.
    A natal kick on the collapsing star may have components in radial, tangential, and polar directions.    
    The binary orbit is prograde (or retrograde) if the component spins of the star and black hole (BH) are aligned (or anti-aligned) with the orbital angular momentum vector.
    }
    \label{fig:cartoon}
\end{figure}

For a circular pre-SN binary \citep[e.g.][]{2014LRR....17....3P,2023pbse.book.....T}, the change in orbital separation and eccentricity following the SN is given as
\begin{equation}\label{eq:separation}
    \frac{a_{\rm post}}{a_{\rm{pre}}} = \Bigg[2 - \Bigg(\frac{M_{\rm pre}}{M_{\rm post}}\Bigg) \Bigg( \frac{v_{\rm r}^2 + v_{\rm p}^2 + (V_{\rm pre}+v_{\rm t})^2}{V_{\rm pre}^2} \Bigg) \Bigg]^{-1}
\end{equation}
and 
\begin{equation}\label{eq:eccentricity}
    1-e^2 = \Bigg(\frac{M_{\rm pre}}{M_{\rm post}}\Bigg) \Bigg(\frac{a_{\rm pre}}{a_{\rm post}}\Bigg) \Bigg(\frac{v_{\rm p}^2+(V_{\rm{pre}}+v_{\rm t})^2}{V_{\rm pre}^2}\Bigg),
\end{equation}
where $a$ is the semi-major axis, $e$ is the eccentricity, $M$ is the total mass, $V$ is the relative velocity (in the tangential direction, see Figure \ref{fig:cartoon}), and the `pre' and `post' subscripts indicate the quantities before and after the SN, respectively.
The natal kick vector, $\vec v$, consists of components in the radial ($v_{\rm r}$), tangential ($v_{\rm t}$), and polar ($v_{\rm p}$) directions within the frame of reference of the exploding star.

In binary systems, there are two limiting scenarios for SNe: spherically symmetric mass loss with no natal kick and natal kicks with negligible mass loss. 
The former scenario has been extensively studied in the literature \citep[e.g.][]{1961BAN....15..265B}.
In the following, I focus on exploring the latter scenario, particularly in the context of binary BH mergers.

\section{Binary BHs formed via direct collapse} 

\begin{figure}
    \centering
    \includegraphics[width=\columnwidth]{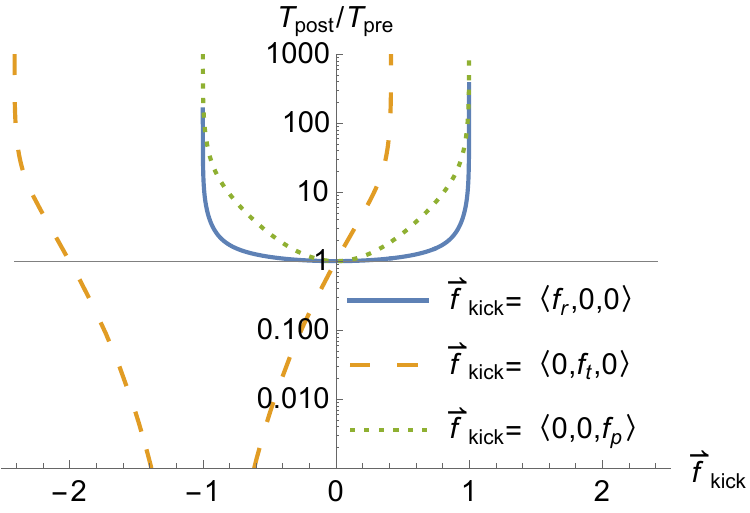}    
    \caption{
    Ratio of the time to coalescence via gravitational-wave emission, parametrised by the orbital properties just after ($T_{\rm post}$) and just before ($T_{\rm pre}$)  binary BH formation, as a function of $\vec f_{\rm kick}:=\vec v/V_{\rm pre}$, where $\vec v$ is the natal kick vector and $V_{\rm pre}$ is the magnitude of the relative orbital velocity before binary BH formation.
    I illustrate three scenarios: natal kicks occurring either in the radial (solid blue), tangential (dashed orange), or polar (dotted green) direction.
    }
    \label{fig:time-to-coalescence}
\end{figure}

In the direct collapse scenario a BH forms without an explosion \citep[e.g.][]{2012ARNPS..62..407J}.
In this scenario, which is often invoked in the context of the formation of BHs more massive than about 10 $M_{\odot}$ \citep[e.g.][and references therein]{2025ApJ...987..164B}, there is negligible baryonic mass ejecta and no associated hydrodynamic natal kick. 
The only radiated energy is approximately $0.1\ M_{\odot}$ via neutrinos, which accounts for less than 1\% of the mass of stellar-mass BHs with masses greater than about $10\ M_{\odot}$.
Therefore, for direct collapse leading to binary BH formation, I assume $M_{\rm pre} = M_{\rm post}$ (i.e. no baryonic mass is ejected) and that any resulting kicks stem predominantly from asymmetries in neutrino emission.
To simplify Eqs. \eqref{eq:separation} and \eqref{eq:eccentricity}, the natal kick components are parametrised as
\begin{equation}
    \vec f_{\rm kick}:=\langle v_{r}, v_{t}, v_{p} \rangle/V_{\rm pre}=\langle f_{\rm r}, f_{\rm t}, f_{\rm p} \rangle.
\end{equation}
This transformation enables me to analyse how natal kicks influence the orbital parameters of newly formed binary BHs. 
The magnitude of (neutrino-induced) natal kicks in BHs remains a topic of active debate in astrophysics, with predictions ranging from roughly 1 km/s \citep[][and references therein]{2024Ap&SS.369...80J} to $\sim$100 km/s \citep[e.g.][]{2024ApJ...963...63B}.

I proceed to address the impact of natal kicks in the time to coalescence of binary BHs.
To do so, I compare the time to coalescence just before ($T_{\rm pre}$) and just after ($T_{\rm post}$) the formation of a binary BH. I\ followed
\cite{1964PhRv..136.1224P} in deriving the time to coalescence for two point masses, $M_1$ and $M_2$, as
\begin{equation}\label{eq:time_to_coalescence}
    T=\frac{5c^5a_0^4}{256 G^3 M_1 M_2 (M_1+M_2)}\times F(e_0),
\end{equation}
where $c$ is the speed of light, $G$ is the Gravitational constant, and the subscript `0' denotes initial values.
The function $F(e_0)$ includes an integral that can be solved analytically \citep{1996NCimB.111..631P}.
Here, I use a computationally efficient analytical approximation (accurate to within 3\%), as presented in \cite{2021RNAAS...5..223M}, 
\begin{equation}\label{eq:Ilya}
    F(e_0) \approx (1 + 0.27e_0^{10} + 0.33e_0^{20} + 0.2e_0^{1000})(1-e_0^2)^{7/2},
\end{equation}
which I  used to compute the role of eccentricity following binary BH formation.
In the direct collapse scenario, since the masses remain approximately unchanged, the time to coalescence will be affected exclusively according to the change in separation and eccentricity.
I can rewrite Eqs. \eqref{eq:separation} and \eqref{eq:eccentricity}, exclusively in terms of the parametrised natal kicks components, as
\begin{equation}
    \frac{a_{\rm post}}{a_{\rm pre}} = \frac{1}{2-[f_{\rm r}^2 + f_{\rm p}^2 + (1+f_{\rm t})^2]}
\end{equation}
and
\begin{equation}
    % e = \sqrt{1+[f_{\rm kick,r}^2 + f_{\rm kick,p}^2 + (1+f_{\rm kick,t})^2 - 2][f_{\rm kick,p}^2 + (1 + f_{\rm kick,t})^2]},
    e = \sqrt{1+[f_{\rm r}^2 + f_{\rm p}^2 + (1+f_{\rm t})^2 - 2][f_{\rm p}^2 + (1 + f_{\rm t})^2]},
\end{equation}
respectively.
Figure \ref{fig:time-to-coalescence} presents the ratio $T_{\rm post}/T_{\rm pre}$ as a function of $\vec f_{\rm kick}$ for three illustrative scenarios where natal kicks occur either in the radial, tangential, or polar direction.
Kicks in the radial ($\vec f_{\rm kick} = \langle f_{\rm r}, 0, 0 \rangle$) and polar ($\vec f_{\rm kick} = \langle 0, 0, f_{\rm p} \rangle$) directions increase the time to coalescence, especially as $f_{\rm kick} \rightarrow 1$.
For radial and polar kicks, if $f_{\rm kick} > 1$, the binary is disrupted resulting in two isolated BHs.
Kicks in the tangential direction ($\vec f_{\rm kick} = \langle 0, f_{\rm t}, 0 \rangle$) are more complex.
A kick along the orbital velocity ($f_{\rm t}>0$) increases the time to coalescence, but disrupts the binary if $f_{\rm t} \geq \sqrt{2}-1$.
Conversely, a kick opposite to the orbital velocity with $-2 < f_{\rm t} <0$ decreases the time to coalescence, with a divergence at $f_{\rm t} \rightarrow -1$ as the orbit becomes parabolic.
For values of $-\sqrt{2}-1 < f_{\rm t} < -2$, the time to coalescence always increases; if $f_{\rm t} < -\sqrt{2}-1$, the binary is disrupted.
A tangential, anti-aligned kick with $f_{\rm t} < -1$ results in a retrograde orbit, in which the component BH spins are anti-aligned with the orbital angular momentum vector (see Figure \ref{fig:cartoon}). 

\begin{figure}
    \centering
     \includegraphics[width=\columnwidth]{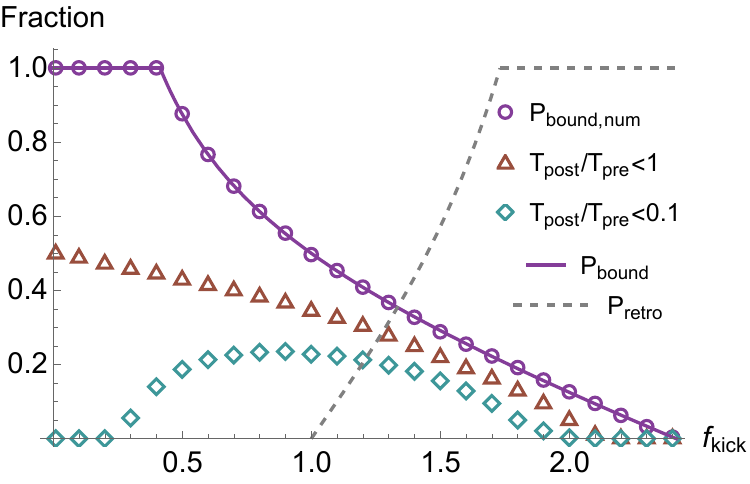}
    \caption{
    Fraction of systems of interest as a function of $|\vec f_{\rm kick}|:= f_{\rm kick}$ under the assumption that the direction of the natal kick is isotropic.
    Bound systems ($P_{\rm bound,num}$) are shown as purple circles.
    Bound binaries in which $T_{\rm post}/T_{\rm pre}<1$ (or $T_{\rm post}/T_{\rm pre}<0.1$) are shown as brown triangles (or teal diamonds).
    Analytic solutions are shown for bound systems ($P_{\rm bound}$, purple solid line) and the fraction of those bound systems that lead to retrograde orbits ($P_{\rm retro}$, gray dashed line).
    }
    \label{fig:fraction-of-systems}
\end{figure}

Figure \ref{fig:fraction-of-systems} illustrates the effect of natal kicks on the time to coalescence. 
For a given magnitude of $f_{\rm kick}$, I sampled $10^6$ isotropically distributed natal kick vectors.
I computed the fraction of systems that remain bound, which can be determined analytically \citep[e.g. Equation 13.4 in][]{2023pbse.book.....T} and is expressed as
\begin{equation}
    P_{\rm bound} = \frac{1}{2} \Bigg( 1+\frac{1-f_{\rm kick}^2}{2f_{\rm kick}} \Bigg), 
\end{equation}
along with the probability that a bound binary results in a retrograde orbit \citep[see Equation 2.20 in][]{1995MNRAS.274..461B}, expressed as
\begin{equation}
    P_{\rm retro} = \frac{1-(1/f_{\rm kick})}{1+(1-f_{\rm kick}^2)/2f_{\rm kick}},
\end{equation}
both of which are bound between 0 and 1.
I  quantified the fraction of systems where the time to coalescence decreases after the formation of the binary BH, denoted as $T_{\rm post}/T_{\rm pre}<1$, and specifically highlight those with a decrease of more than an order of magnitude, $T_{\rm post}/T_{\rm pre}<0.1$.
For $f_{\rm kick} \lesssim 0.2$, all systems remain bound; although about half result in a decreased time to coalescence, none change drastically.
In the range $0.2 \lesssim f_{\rm kick} < \sqrt{2}-1$, all systems remain bound and nearly one-fifth see their time to coalescence decreased by more than an order of magnitude. 
For $f_{\rm kick} > \sqrt{2}-1$, the fraction of bound systems consistently drops.
The fraction of systems with a substantial decrease in time to coalescence peaks at approximately 0.23 for $f_{\rm kick} \approx 1$, then decreases for $f_{\rm kick}>1$.
Bound orbits with $f_{\rm kick}>1$ tend to have retrograde orbits \citep[cf. Figure 3 in ][]{1995MNRAS.274..461B}.

Broadly, there are three regimes relevant in this context, as detailed below: 
\begin{enumerate}
    \item 
    Low-magnitude natal kicks ($f_{\rm kick}\lesssim 0.2$) which always result in a bound binary with negligible impact on their time to coalescence. 
    \item 
    Natal kicks with magnitudes comparable to the orbital velocity before the formation of the binary BH ($f_{\rm kick}\approx 1$) can significantly reduce the time to coalescence, but are also likely to disrupt the binary. 
    \item 
    Large-magnitude natal kicks ($f_{\rm kick}>1$) rarely result in a bound binary. However, when they do, they often lead to retrograde orbits and the time to coalescence may decrease substantially. 
    These potentially retrograde orbits have been initially discussed in models of X-ray binaries \citep{1995MNRAS.274..461B} and, more recently, in the context of binary BH mergers \citep{2024arXiv241203461B}.
\end{enumerate}

I\ note that my analysis omits the following: i) the pre-SN binary may be eccentric \citep[see appendices of][for SNe in eccentric binaries]{2002MNRAS.329..897H,2002ApJ...573..283P}; ii) the explosion might not be instantaneous \cite[see][and references therein]{2024ApJ...972L..18H}; and iii) there could be some mass loss, which tends to widen the orbit and increase its eccentricity while keeping the pericenter constant.

\section{Characteristic orbit of the pre-SN binary}
\begin{figure}
    \centering
    \includegraphics[width=\columnwidth]{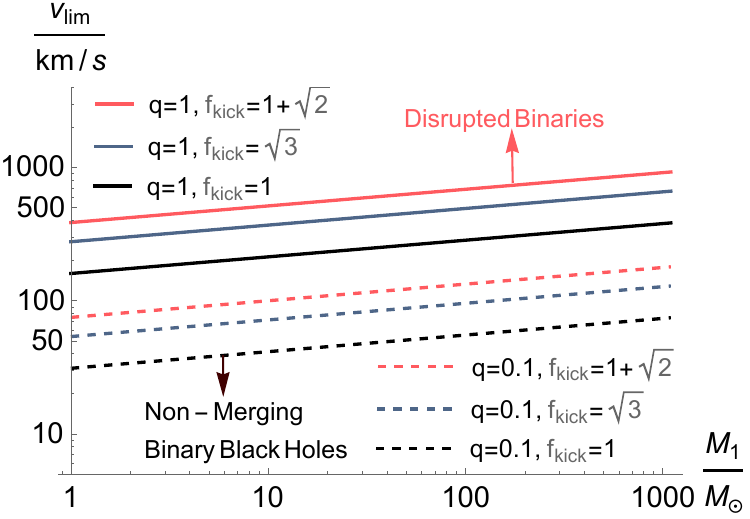}
    \caption{
    Limit orbital speed ($v_{\lim}$) as a function of mass for a star that is about to experience complete collapse in a binary system. 
    The limit orbital speed determines the minimum relative orbital speed a star in a binary must have in order to merge within the age of the Universe (black lines). 
    For a given binary system, a natal kick $f_{\rm kick} \geq \sqrt{3}$ will lead to a retrograde orbit (blue lines) and a natal kick above $f_{\rm kick} \geq 1+\sqrt{2}$ will disrupt the binary (red lines).
    The characteristic speed depends on the mass ratio ($q:=M_2/M_1$), which I consider to be either $q=0.1$ (dashed lines) or $q=1$ (solid lines).
    }
    \label{fig:characteristic_speed}
\end{figure}

The results presented thus far are scale-free. For binary BH mergers, one characteristic scale is determined by the maximum separation at which they would merge within the age of the Universe ($ a_{\rm lim}$), which can be estimated using Eq. \eqref{eq:time_to_coalescence}. 
Subsequently, this allows for the determination of a limit orbital speed,
\begin{equation}
    v_{\rm lim}=\frac{M_{\rm BH}}{M_{\rm pre}}\sqrt{\frac{GM_{\rm pre}}{a_{\rm lim}}},
\end{equation}
which represents the speed of the star, just before the formation of the second BH, in its frame of reference (Figure \ref{fig:cartoon}).

Figure \ref{fig:characteristic_speed} presents $v_{\rm lim}$ for binaries with primary star mass between $1 \leq M_1/M_{\odot} \leq 1000$ and mass ratios, $q:=M_2/M_1=\{0.1,1\}$. 
The black line indicates the $v_{\rm lim}$ threshold that determines which binary BHs (formed without natal kicks) can merge within the age of the Universe; values below this line correspond to binaries that are too wide to merge and therefore cannot become gravitational-wave sources.
For a given mass ratio, the region between the blue and red lines corresponds to natal kick magnitudes, which (for that specific orbital configuration) always result in retrograde orbits. 
Finally, the red line marks the limit above which binaries become gravitationally unbound.

For a star mass of $10~M_{\odot}$ with a mass ratio $q=0.1$ (or $q=1$), the limit orbital speed is $v_{\rm lim}\approx 41$ km/s (or $v_{\rm lim}\approx 213$ km/s).
This implies that to disrupt the binary, the natal kicks must exceed roughly $100$ km/s (or $515$ km/s).
It is important to note that progenitors of binary BH mergers may exhibit higher orbital speeds, which would in turn raise these limit values. Therefore, our current estimates for the limiting orbital speeds can be considered conservative.
In the context of the complete collapse scenario, most models suggest that neutrino natal kicks are $\sim 1-10$ km/s, with values reaching a few tens of km/s considered significant \citep{2024Ap&SS.369...80J}.
Even models with large neutrino emission asymmetries cannot account for natal kicks of $\gtrsim 100$ km/s \citep{{2024ApJ...963...63B}}.
Based on a single observation of the X-ray-quiescent BH binary VFTS 243, the neutrino natal kick has been estimated to be around $0-10\ \rm{km/s}$ \citep{2024PhRvL.132s1403V}, consistent with a result of no natal kick.
Thus, if our current understanding of neutrino natal kicks is correct, they would have a negligible impact on the binary BH merger time and are highly unlikely to result in a retrograde orbit.

Within the Local Group, several observations have identified high-mass stars and BH binaries at intermediate stages on their path to becoming binary BHs. 
Notably, the systems Cygnus X-1 \citep{2003Sci...300.1119M} and VFTS 243 \citep{2022NatAs...6.1085S} feature BHs that are candidates for complete collapse. 
In each case, the  relative orbital velocity of the BH exceeds 100 km/s. 
If these velocities are representative of similar systems, especially just before the formation of the BH, only substantial natal kicks, most likely hydrodynamic in nature, could yield orbital flips.
Although there is currently no consensus on the magnitude of BH natal kicks, tentative kinematic evidence suggests that at least some BHs receive kicks above 100 km/s \citep{2025PASP..137c4203N}.

I briefly outline one scenario where neutrino natal kicks might influence BHs with stellar companions. 
Models of isolated binary evolution predict a large population of BHs paired mainly with low-mass companions in the Galaxy \citep{2017ApJ...850L..13B,2020A&A...638A..94O}. 
Many of these systems are expected to originate in wide binaries \citep{2024A&A...692A.141K}, making them sensitive to both low and moderate neutrino natal kicks. 
These wide binaries are likely to be initially eccentric and their spins need not be preferentially aligned, which makes probing orbital flips challenging. 
Nonetheless, one potential effect is that some wide binaries could be perturbed into tighter, more eccentric orbits, thereby becoming observable by Gaia \citep{2016A&A...595A...1G}. 
While disentangling the influence of individual wide BH binaries might be difficult, their collective impact could be manifested at specific separations; this could support, for example, a super-thermal distribution.

\section{Implication for spin-orbit alignment}
Finally, natal kicks have often been considered to affect BH spins:\ a quantity that is indirectly inferred from gravitational-wave observations of binary BH mergers. 
In gravitational-wave astronomy, the best-measured spin quantity of binary BHs is given by the effective spin parameter
\begin{equation}
    \chi_{\rm eff}=\frac{\chi_1\cos{\theta_1}+q\chi_2\cos{\theta_2}}{1+q},
\end{equation}
where $\chi_{1,2}$ is the magnitude of the dimensionless spin vector, $\cos{\theta_{1,2}}$ is the misalignment angle between the orbital angular momentum vector and the respective spin vectors, and $q=M_2/M_1\leq 1$ is the mass ratio \citep[e.g.][and references therein]{2021ApJ...921L..15G}. 
The effective spin parameter ranges between $-1 \leq \chi_{\rm eff} \leq +1$.
It reaches –1 (or +1) for maximally spinning BHs that are completely anti-aligned (or aligned) with the orbital angular momentum vector.
It tends to be close to zero when the spins are small or cancel each other out.
For binary BHs formed from isolated binary stars, it has often been assumed that $\chi_{\rm eff} \gtrsim 0$, rarely (if ever) leading to negative values.
However, \cite{2022ApJ...938...66T} showed that randomizing or tossing BH spin axes can explain the observed distribution of the effective spin, alleviating the need for strong kicks.
Similarly, \cite{2024arXiv241203461B} finds that relaxing the assumption of spin alignment prior to the second SN can lead to precessing and retrograde binary BH mergers in isolated environments.

In our model, there are many caveats for a binary to lead to a retrograde orbit.
Specifically, at the moment of the second collapse leading to the formation of the binary BH: 
\begin{enumerate}[i]
    \item the spins must be non-zero and aligned; 
    \item there must be no spin tossing during BH formation (i.e. the spin direction must not be random);
    \item the natal kick needs to be either moderate or large; this is something that I show is particularly difficult to reconcile with the complete collapse scenario;  
    \item finally, the direction of the natal kick need to be either isotropically distributed or biased toward the negative tangential direction.
\end{enumerate}
If the above conditions are met, the binary BH would yield $\chi_{\rm eff}<0$. This is a value often linked to a dynamical formation scenario and the time to coalescence could be considerably reduced. 
Our findings suggest that neutrino natal kicks are unlikely to induce orbital flips. 
If orbital flips do occur, they are more plausibly explained by (larger) hydrodynamic kicks or spin-tossing.
Consequently, inferring BH natal kicks from gravitational-wave observations may be challenging. 
The best prospects could come from electromagnetic observations within the Local Group; specifically, future discoveries of BHs in massive binary systems similar to Cygnus X-1 or VFTS 243. 
In addition, BHs with luminous companions in wide binaries (potentially detectable by Gaia) could offer valuable insights into BH formation.

\begin{acknowledgements}
I thank Zhanwen Han, Aldana Grichener, Ryosuke Hirai, Hans-Thomas Janka, Jakub Klencki, Mathieu Renzo, Edwin Santiago-Leandro, Taeho Ryu, and Lorenz Zwick for useful discussions and helpful suggestions.
I thank Ilya Mandel, Jakob Stegmann, and Thomas M. Tauris for a careful reading of the manuscript.
I thank the anonymous referee for their useful suggestions which improved the manuscript.
\end{acknowledgements}

\bibliographystyle{aa} % style aa.bst
\bibliography{references} % your references Yourfile.bib

\begin{thebibliography}{31}
\expandafter\ifx\csname natexlab\endcsname\relax\def\natexlab#1{#1}\fi

\bibitem[{{Andrews} \& {Kalogera}(2022)}]{2022ApJ...930..159A}
{Andrews}, J.~J. \& {Kalogera}, V. 2022, \apj, 930, 159

\bibitem[{{Baibhav} \& {Kalogera}(2024)}]{2024arXiv241203461B}
{Baibhav}, V. \& {Kalogera}, V. 2024, arXiv e-prints, arXiv:2412.03461

\bibitem[{{Beniamini} \& {Piran}(2024)}]{2024ApJ...966...17B}
{Beniamini}, P. \& {Piran}, T. 2024, \apj, 966, 17

\bibitem[{{Blaauw}(1961)}]{1961BAN....15..265B}
{Blaauw}, A. 1961, \bain, 15, 265

\bibitem[{{Brandt} \& {Podsiadlowski}(1995)}]{1995MNRAS.274..461B}
{Brandt}, N. \& {Podsiadlowski}, P. 1995, \mnras, 274, 461

\bibitem[{{Breivik} {et~al.}(2017){Breivik}, {Chatterjee}, \&
  {Larson}}]{2017ApJ...850L..13B}
{Breivik}, K., {Chatterjee}, S., \& {Larson}, S.~L. 2017, \apjl, 850, L13

\bibitem[{{Burrows} {et~al.}(2025){Burrows}, {Wang}, \&
  {Vartanyan}}]{2025ApJ...987..164B}
{Burrows}, A., {Wang}, T., \& {Vartanyan}, D. 2025, \apj, 987, 164

\bibitem[{{Burrows} {et~al.}(2024){Burrows}, {Wang}, {Vartanyan}, \&
  {Coleman}}]{2024ApJ...963...63B}
{Burrows}, A., {Wang}, T., {Vartanyan}, D., \& {Coleman}, M. S.~B. 2024, \apj,
  963, 63

\bibitem[{{Gaia Collaboration} {et~al.}(2016){Gaia Collaboration}, {Prusti},
  {de Bruijne}, {Brown}, {Vallenari}, {Babusiaux}, {Bailer-Jones}, {Bastian},
  {Biermann}, {Evans}, {Eyer}, {Jansen}, {Jordi}, {Klioner}, {Lammers},
  {Lindegren}, {Luri}, {Mignard}, {Milligan}, {Panem}, {Poinsignon},
  {Pourbaix}, {Randich}, {Sarri}, {Sartoretti}, {Siddiqui}, {Soubiran},
  {Valette}, {van Leeuwen}, {Walton}, {Aerts}, {Arenou}, {Cropper}, {Drimmel},
  {H{\o}g}, {Katz}, {Lattanzi}, {O'Mullane}, {Grebel}, {Holland}, {Huc},
  {Passot}, {Bramante}, {Cacciari}, {Casta{\~n}eda}, {Chaoul}, {Cheek}, {De
  Angeli}, {Fabricius}, {Guerra}, {Hern{\'a}ndez}, {Jean-Antoine-Piccolo},
  {Masana}, {Messineo}, {Mowlavi}, {Nienartowicz}, {Ord{\'o}{\~n}ez-Blanco},
  {Panuzzo}, {Portell}, {Richards}, {Riello}, {Seabroke}, {Tanga},
  {Th{\'e}venin}, {Torra}, {Els}, {Gracia-Abril}, {Comoretto},
  {Garcia-Reinaldos}, {Lock}, {Mercier}, {Altmann}, {Andrae}, {Astraatmadja},
  {Bellas-Velidis}, {Benson}, {Berthier}, {Blomme}, {Busso}, {Carry},
  {Cellino}, {Clementini}, {Cowell}, {Creevey}, {Cuypers}, {Davidson}, {De
  Ridder}, {de Torres}, {Delchambre}, {Dell'Oro}, {Ducourant}, {Fr{\'e}mat},
  {Garc{\'\i}a-Torres}, {Gosset}, {Halbwachs}, {Hambly}, {Harrison}, {Hauser},
  {Hestroffer}, {Hodgkin}, {Huckle}, {Hutton}, {Jasniewicz}, {Jordan},
  {Kontizas}, {Korn}, {Lanzafame}, {Manteiga}, {Moitinho}, {Muinonen},
  {Osinde}, {Pancino}, {Pauwels}, {Petit}, {Recio-Blanco}, {Robin}, {Sarro},
  {Siopis}, {Smith}, {Smith}, {Sozzetti}, {Thuillot}, {van Reeven}, {Viala},
  {Abbas}, {Abreu Aramburu}, {Accart}, {Aguado}, {Allan}, {Allasia},
  {Altavilla}, {{\'A}lvarez}, {Alves}, {Anderson}, {Andrei}, {Anglada Varela},
  {Antiche}, {Antoja}, {Ant{\'o}n}, {Arcay}, {Atzei}, {Ayache}, {Bach},
  {Baker}, {Balaguer-N{\'u}{\~n}ez}, {Barache}, {Barata}, {Barbier}, {Barblan},
  {Baroni}, {Barrado y Navascu{\'e}s}, {Barros}, {Barstow}, {Becciani},
  {Bellazzini}, {Bellei}, {Bello Garc{\'\i}a}, {Belokurov}, {Bendjoya},
  {Berihuete}, {Bianchi}, {Bienaym{\'e}}, {Billebaud}, {Blagorodnova},
  {Blanco-Cuaresma}, {Boch}, {Bombrun}, {Borrachero}, {Bouquillon}, {Bourda},
  {Bouy}, {Bragaglia}, {Breddels}, {Brouillet}, {Br{\"u}semeister},
  {Bucciarelli}, {Budnik}, {Burgess}, {Burgon}, {Burlacu}, {Busonero}, {Buzzi},
  {Caffau}, {Cambras}, {Campbell}, {Cancelliere}, {Cantat-Gaudin}, {Carlucci},
  {Carrasco}, {Castellani}, {Charlot}, {Charnas}, {Charvet}, {Chassat},
  {Chiavassa}, {Clotet}, {Cocozza}, {Collins}, {Collins}, \&
  {Costigan}}]{2016A&A...595A...1G}
{Gaia Collaboration}, {Prusti}, T., {de Bruijne}, J.~H.~J., {et~al.} 2016,
  \aap, 595, A1

\bibitem[{{Galaudage} {et~al.}(2021){Galaudage}, {Talbot}, {Nagar}, {Jain},
  {Thrane}, \& {Mandel}}]{2021ApJ...921L..15G}
{Galaudage}, S., {Talbot}, C., {Nagar}, T., {et~al.} 2021, \apjl, 921, L15

\bibitem[{{Hirai} {et~al.}(2024){Hirai}, {Podsiadlowski}, {Heger}, \&
  {Nagakura}}]{2024ApJ...972L..18H}
{Hirai}, R., {Podsiadlowski}, P., {Heger}, A., \& {Nagakura}, H. 2024, \apjl,
  972, L18

\bibitem[{{Hurley} {et~al.}(2002){Hurley}, {Tout}, \&
  {Pols}}]{2002MNRAS.329..897H}
{Hurley}, J.~R., {Tout}, C.~A., \& {Pols}, O.~R. 2002, \mnras, 329, 897

\bibitem[{{Janka}(2012)}]{2012ARNPS..62..407J}
{Janka}, H.-T. 2012, Annual Review of Nuclear and Particle Science, 62, 407

\bibitem[{{Janka} \& {Kresse}(2024)}]{2024Ap&SS.369...80J}
{Janka}, H.-T. \& {Kresse}, D. 2024, \apss, 369, 80

\bibitem[{{Kalogera}(1996)}]{1996ApJ...471..352K}
{Kalogera}, V. 1996, \apj, 471, 352

\bibitem[{{Kalogera}(2000)}]{2000ApJ...541..319K}
{Kalogera}, V. 2000, \apj, 541, 319

\bibitem[{{Kruckow} {et~al.}(2024){Kruckow}, {Andrews}, {Fragos}, {Holl},
  {Bavera}, {Briel}, {Gossage}, {Kovlakas}, {Rocha}, {Sun}, {Srivastava},
  {Xing}, \& {Zapartas}}]{2024A&A...692A.141K}
{Kruckow}, M.~U., {Andrews}, J.~J., {Fragos}, T., {et~al.} 2024, \aap, 692,
  A141

\bibitem[{{Mandel}(2021)}]{2021RNAAS...5..223M}
{Mandel}, I. 2021, Research Notes of the American Astronomical Society, 5, 223

\bibitem[{{Mirabel} \& {Rodrigues}(2003)}]{2003Sci...300.1119M}
{Mirabel}, I.~F. \& {Rodrigues}, I. 2003, Science, 300, 1119

\bibitem[{{Nagarajan} \& {El-Badry}(2025)}]{2025PASP..137c4203N}
{Nagarajan}, P. \& {El-Badry}, K. 2025, \pasp, 137, 034203

\bibitem[{{Olejak} {et~al.}(2020){Olejak}, {Belczynski}, {Bulik}, \&
  {Sobolewska}}]{2020A&A...638A..94O}
{Olejak}, A., {Belczynski}, K., {Bulik}, T., \& {Sobolewska}, M. 2020, \aap,
  638, A94

\bibitem[{{Peters}(1964)}]{1964PhRv..136.1224P}
{Peters}, P.~C. 1964, Physical Review, 136, 1224

\bibitem[{{Pfahl} {et~al.}(2002){Pfahl}, {Rappaport}, \&
  {Podsiadlowski}}]{2002ApJ...573..283P}
{Pfahl}, E., {Rappaport}, S., \& {Podsiadlowski}, P. 2002, \apj, 573, 283

\bibitem[{{Pierro} \& {Pinto}(1996)}]{1996NCimB.111..631P}
{Pierro}, V. \& {Pinto}, I.~M. 1996, Nuovo Cimento B Serie, 111B, 631

\bibitem[{{Postnov} \& {Yungelson}(2014)}]{2014LRR....17....3P}
{Postnov}, K.~A. \& {Yungelson}, L.~R. 2014, Living Reviews in Relativity, 17,
  3

\bibitem[{{Shenar} {et~al.}(2022){Shenar}, {Sana}, {Mahy}, {El-Badry},
  {Marchant}, {Langer}, {Hawcroft}, {Fabry}, {Sen}, {Almeida}, {Abdul-Masih},
  {Bodensteiner}, {Crowther}, {Gieles}, {Gromadzki}, {H{\'e}nault-Brunet},
  {Herrero}, {de Koter}, {Iwanek}, {Koz{\l}owski}, {Lennon}, {Ma{\'\i}z
  Apell{\'a}niz}, {Mr{\'o}z}, {Moffat}, {Picco}, {Pietrukowicz}, {Poleski},
  {Rybicki}, {Schneider}, {Skowron}, {Skowron}, {Soszy{\'n}ski},
  {Szyma{\'n}ski}, {Toonen}, {Udalski}, {Ulaczyk}, {Vink}, \&
  {Wrona}}]{2022NatAs...6.1085S}
{Shenar}, T., {Sana}, H., {Mahy}, L., {et~al.} 2022, Nature Astronomy, 6, 1085

\bibitem[{{Tauris}(2022)}]{2022ApJ...938...66T}
{Tauris}, T.~M. 2022, \apj, 938, 66

\bibitem[{{Tauris} {et~al.}(2017){Tauris}, {Kramer}, {Freire}, {Wex}, {Janka},
  {Langer}, {Podsiadlowski}, {Bozzo}, {Chaty}, {Kruckow}, {van den Heuvel},
  {Antoniadis}, {Breton}, \& {Champion}}]{2017ApJ...846..170T}
{Tauris}, T.~M., {Kramer}, M., {Freire}, P.~C.~C., {et~al.} 2017, \apj, 846,
  170

\bibitem[{{Tauris} \& {Takens}(1998)}]{1998A&A...330.1047T}
{Tauris}, T.~M. \& {Takens}, R.~J. 1998, \aap, 330, 1047

\bibitem[{{Tauris} \& {van den Heuvel}(2023)}]{2023pbse.book.....T}
{Tauris}, T.~M. \& {van den Heuvel}, E. P.~J. 2023, {Physics of Binary Star
  Evolution. From Stars to X-ray Binaries and Gravitational Wave Sources}

\bibitem[{{Vigna-G{\'o}mez} {et~al.}(2024){Vigna-G{\'o}mez}, {Willcox},
  {Tamborra}, {Mandel}, {Renzo}, {Wagg}, {Janka}, {Kresse}, {Bodensteiner},
  {Shenar}, \& {Tauris}}]{2024PhRvL.132s1403V}
{Vigna-G{\'o}mez}, A., {Willcox}, R., {Tamborra}, I., {et~al.} 2024, \prl, 132,
  191403

\end{thebibliography}
\end{document}